arXiv:hep-ph/0204019 2 Apr 2002

# Bounds on the W-boson Electric Dipole Moment – Revisited

P. Saxena[*], Prachi Parashar[♣], N.K. Sharma[•], Ashok K. Nagawat[♠] and Sardar Singh[©]

Department of Physics, University of Rajasthan, Jaipur (India)– 302 004.

## Abstract

Keeping in view the suggestion [7] that the cut-off procedure involves a good deal of uncertainity in the prediction of E.D.M. of W – boson, we have re-examined the earlier calculation by Marciano and Queijeiro [2] by replacing the Cut-off regularization process by BPHZ regularization [8]. This works in a clean and unambiguous manner without involving any approximation. We also examine apparently inapplicable approximations like using $m_f = m_{f'}$ in [2]. The bounds on $\lambda_W$ and $d_W$ are significantly changed in all cases as compared with those reported in [2]. The necessary caution that is to be exercised while using approximations is explicitly pointed out.

[*]pranav_saxena19@hotmail.com
[♣]prachi_parashar@hotmail.com
[•]nksharma_thep@hotmail.com
[♠]nagawat@hotmail.com
[©]singhjaipur@hotmail.com

## I. INTRODUCTION

In the past few years, there have been some efforts to evaluate Electric Dipole Moment (EDM) of W- boson [1-3]. In 1986, Marciano and Queijeiro [2] obtained limit on $\lambda_W$ and $d_W$ of W- boson updating an earlier suggestion by Salzman and Salzman [1]. They made use of the concept of induced fermion electric dipole moment, which is induced in the presence of W- boson or vice-versa. In order to overcome the ultraviolet divergence, they made use of the cut-off procedure, first used by Paulli-Villars [4]. This work has been extensively been used by the later authors in this field in their calculations [3,5,6]. As rightly pointed out by Barr and Marciano [7], the bounds derived from the cut-off dependent loop effects can sometimes be misleading. Unforeseen cancellations could reduce the value of EDM estimate particularly if the scale of new physics defined through the cut-off dependent parameter $\Lambda$ is equal to $m_W$. Apart from this, the authors of ref. [2] made use of some approximations, which might have eventually influenced the outcome. In particular, the assumption making $m_f = m_{f'}$, is not physically correct. In order to check whether the observation made in [7] plus the approximations used in [2] have some impact on the results we have re-performed this calculation. In order to overcome the uncertainty in the use of cut-off procedure, we make use of the versatile BPHZ regularization [8] procedure without making any approximation whatsoever. The calculation for $D_\mu$, $\lambda_W$ and $d_W$ go on very smoothly till the end. However, the limits obtained on $\lambda_W$ and $d_W$ are drastically different from those given in ref. [2].

Next to check the effects of the approximation made in [2], we have re-performed complete calculation for $D_\mu$, $\lambda_W$ and $d_W$ using Pauli-Villars cut-off method [4] without making any approximation. Here again the limits obtained on $\lambda_W$ and $d_W$ are significantly changed as compared with those reported in ref. [2]. Thus this outcome apart from substantiating the observation made by Barr and Marciano [7], also points towards being careful about making approximations while carrying out renormalization program. This aspect will be further elucidated below.

## II. CALCULATIONS USING BPHZ REGULARIZATION PROCEDURE

Beginning with the CP- violating amplitude reported in ref. [2], namely, Eq. (5) of [2] corresponding to fig. (1), we notice that the integral is logarithmically divergent. Therefore BPHZ regularization scheme [8] can be justifiably applied [9]. Using this scheme and after a lengthy and tedious algebra we obtain for the CP- violating amplitude, the expression:

$$D_\mu = -\frac{\imath e g^2}{128\pi^2}\frac{\lambda_W}{m_W^2}\bar{u}(p_2)q^\nu\left(\not{p}_2\sigma_{\mu\nu}-\sigma_{\mu\nu}\not{p}_1\right)\chi\left(\frac{\gamma_5-1}{2}\right)u(p_1); \tag{1a}$$

which leads to

$$d_f = \frac{eG_F m_f \lambda_W}{8\sqrt{2}\pi^2}\chi,$$

$$= \left(4.126\times10^{-21}ecm\right)\frac{m_f\lambda_W}{2GeV}\chi, \tag{1b}$$

where

$$\chi = \left[\left\{\frac{R'^3-R'^2R-R'R^2-2R'R+R^3-R'^2-R'-R^2-R+1}{R^2\sqrt{R'^2-2R'R-2R'+R^2-2R+1}}\right\}\right.$$

$$\times\ln\left\{\frac{(R'-R+1)+\sqrt{R'^2-2R'R-2R'+R^2-2R+1}}{(R'-R+1)-\sqrt{R'^2-2R'R-2R'+R^2-2R+1}}\right\}-\left(\frac{R'^2+R^2-1}{R^2}\right)\ln(R')+$$

$$\left(\frac{2R'R^2+6R'R-R^3+10R^2+6R}{3R^2}\right)\Bigg]. \tag{1c}$$

with $R=\dfrac{m_f^2}{m_W^2}$, $R'=\dfrac{m_{f'}^2}{m_W^2}$.

In arriving at Eq. (1), the following identities have also been used:

$$\frac{\imath}{2}\varepsilon_{\alpha\beta\mu\nu}\,\sigma^{\alpha\beta}=\sigma_{\mu\nu}\gamma_5\,,\qquad \varepsilon_{\alpha\beta\mu\nu}\,\gamma^{\alpha}=\frac{1}{2}\left(\gamma_\nu\sigma_{\beta\mu}+\sigma_{\beta\mu}\gamma_\nu\right)\gamma_5\,; \tag{2a}$$

$$\varepsilon_{\alpha\beta\mu\nu}\,p_1^\alpha\,p_1^\beta=\varepsilon_{\alpha\beta\mu\nu}\,p_2^\alpha\,p_2^\beta=\varepsilon_{\alpha\beta\mu\nu}q^\nu q^\alpha q^\beta=0. \tag{2b}$$

It may be emphasized that we have used $m_f\neq m_{f'}$ and $R,R'\neq 0$ in this calculation. We notice that the factor $\left[\ln\dfrac{\Lambda^2}{m_W^2}+O(1)\right]$ of ref. [2] Eq. (8) is replaced by $\chi$ in our formulation. We have evaluated $\chi$ and bounds on $\lambda_W$ and $d_W$ by using the experimental bounds [10] on $d_e$, $d_\mu$, $d_\tau$ and $d_n$. The results are shown in Table 1. For comparison we have shown the corresponding values arising from the calculations of ref. [2], Eq. (9) in Table 2 where the aforesaid approximations have been used.

## III. CALCULATIONS USING CUT-OFF PROCEDURE ($m_f\neq m_{f'}, R, R'\neq 0$)

Again beginning with the Eq. (5) of ref. [2] we obtained after a very lengthy algebraic manipulation the following form of expression for $D_\mu$:

$$D_\mu=\frac{-\imath e g^2}{64\pi^2 m_W^2}\lambda_W\,\bar{u}(p_2)q^\nu\left[p_2\sigma_{\mu\nu}\gamma_5+\sigma_{\mu\nu}\gamma_5 p_1\right]u(p_1)\times\Re \tag{3}$$

$$d_f=\frac{eG_F}{4\sqrt{2}\pi^2}\lambda_W m_f\Re\,, \tag{4}$$

where

$$\Re = \frac{1}{12(D-1)R^2}\Big[-2R(D-1)(D-5-2R'+2R)+\big\{6D-3D^2+D^3-6R'+3R'^2-3DR'^2+2R'^3$$
$$-6(1+R')R+3R^2(D-1+2R')-2R^3\big\}\ln(D)-(D-1)^3\ln(R')-$$

$$\frac{1}{\sqrt{(1-R'+R)^2-4R}}\Big\{-2R'^4+R'^3(1-8R)+R'^2(9+R-12R^2)+3D\Big\langle R'^3-R'^2(1+R)+(R-1)^2(R+1)$$
$$-R'(R+1)^2\Big\rangle-(R-1)^2(5R+2R^2-1)+R'(8R^3+R^2+10R-7)\Big\}\times$$
$$\ln\left(\frac{1+R'^2-2RR'-2R+R^2+(R'-R+1)\sqrt{(1-R'+R)^2-4R}}{2R'}\right)-$$

$$\frac{1}{\sqrt{(R'-R+D)^2-4RD}}\Big\{4R^2\big\langle-3+D(-3+2D)\big\rangle-(D-R'+R)^2\Big\langle 6+D^2-3R'+D(-3+R'-R)-$$
$$-2(R'-R)^2+3R\Big\rangle-2R\Big\langle 6(D-1)D-6(D+1)^2(D-R'+R)+(5D-3)(D-R'+R)^2\Big\rangle\Big\}\times$$
$$\ln\left(\frac{D^2+R^2+R'^2-2RR'-2RD+(D+R'-R)\sqrt{(D-R'+R)^2\,4RD}}{2R'D}\right)\Big]. \qquad (5)$$

It may be pointed out that the following Feynman parameterization and integrations have been used in the aforesaid calculation:

$$\frac{1}{abcde}=\Gamma(5)\int_0^1 dx_1\int_0^{x_1}dx_2\int_0^{x_2}dx_3\int_0^{x_3}dx_4\,\frac{1}{\left[(R'-R)x_4+Rx_4^2+(x_2-x_4)D-x_2+1\right]^5} \qquad (6)$$

where $x_1$, $x_2$ …… $x_4$ are the Feynman parameters and

$$a=(k^2-k.q-m_W^2)\,,\ b=(k^2+k.q-m_W^2)\,;$$

$$c=(k^2-k.q-\Lambda^2)\,,\ d=(k^2+k.q-\Lambda^2)\,; \qquad (7)$$

$$e=\left(k^2+k.(p_1+p_2)+m_f^2-m_{f'}^2\right)\ \text{and}\ D=\left(\frac{\Lambda^2}{m_W^2}\right).$$

Now if we introduce the approximation $m_f = m_{f'}$ but not $R, R' = 0$ in Eq. (5), the above expression reduces to

$$\Re_{R=R'} = \frac{1}{12(D-1)R^2}\Big\{-2(5R-6DR+D^2R)+(6D-3D^2+D^3-12R)\ \ln(D)-(D-1)^3\ln(R)$$

$$-\frac{1}{\sqrt{1-4R}}\left\langle 1+3D-14R-6DR+28R^2-12DR^2\right\rangle \ln\left(\frac{1-2R+\sqrt{1-4R}}{2R}\right)-$$

$$\frac{1}{\sqrt{D^2-4RD}}\left\langle -6D^2+3D^3-D^4+24DR-6D^2R+2D^3R-12R^2-12DR^2+8D^2R^2\right\rangle$$

$$\ln\left(\frac{D-2R+\sqrt{D^2-4RD}}{2R}\right)\Big\}. \qquad (8)$$

We notice that the approximations $R, R' \to 0$ cannot be used in Eq. (5) and Eq. (8) because both expressions diverge.

The authors of ref. [2] could use these approximations as they have done so before performing Feynman parametric integrations, whereas we are trying to use these after performing the integration over parametric space. To our knowledge, the use of approximations before doing integration in parametric space as is done in ref. [2] need some caution as elucidated below [11]:

The justification of the approximations $R, R' = 0$ used in ref. [2], lies in the fact that these are negligible as compared with the Cut-off parameter $\Lambda$ under the assumption $\Lambda >>> m_W^2$ or $\Lambda \to \infty$. An application of this approximation in the denominator of our Eq. (6) allows only the term $(x_2 - x_4)\frac{\Lambda^2}{m_W^2}$ to survive. This term on being integrated over $x_2$ and $x_4$ diverges when $\Lambda \to \infty$. As such the use of this approximation before integration over parametric space is not justified [11] as is done in ref. [2]. On the other hand there is absolutely no justification to use the approximation $m_f = m_{f'}$ anywhere in the calculation. As such no approximation of the type used in [2] is justifiably applied in the Cut-off procedure calculation.

As the expression (4,5) is not a cozy, as Eq. (8) of [2], we can extract information about $\lambda_W$ and $d_W$ from it by plotting $d_W$ against $\Lambda$ as shown in Fig. (2) using experimental limits on $d_f$ ($f$=$e,\mu,\tau$ and $n$) from [10]. We notice from Fig. (2) that at about $\Lambda$ = *3 TeV*, the values of various $d_W$ are relatively stabilized. These are given in Table 3. The corresponding variation of $d_W$ against $\Lambda$, using Eq. (9) of [2], are shown in Fig. (3). For completeness, we also show the variations of $d_W$ against $\Lambda$ corresponding to our Eq. (4,8) in Fig. (4). The values of $d_W$ corresponding to $\Lambda$ = *3 TeV* using Fig. (3) and (4) are shown in Table 3 and 4 respectively.

As a corollary, we take the limit on $d_W$ corresponding to most stringent limit on neutron EDM ($d_n$) as obtained in our BPHZ regularization procedure as a reference value and obtain there from the limits on $d_e$, $d_\mu$ and $d_\tau$. These are given in Table 5. For another similar exercise, we take $d_W \cong 10^{-30}\, ecm$ and re-calculate $d_e$, $d_\mu$ and $d_\tau$ and $d_n$. We have preferred this value since identical value occurs in the calculation by Booth [5] i.e. $d_W \cong 8\times 10^{-30}\, ecm$, who make use of QCD radiative correction for its evaluation as also in our calculations, namely, $d_W \le 4.228\times 10^{-30}\, ecm$ corresponding the $d_e$ limit (exp.) using BPHZ method (Table 1). The results are shown in Table 6 These values are very close to the corresponding experimental limits. This may lead us to conclude that limits on $d_W$ may lie in the vicinity of $10^{-30}\,ecm$, if these methods are to be believed.

## IV. CONCLUSION

In the absence of any experimental limit on $d_W$ it is not possible to make any definite statement about the outcome of the aforesaid calculations and limits. In reality this is not the objective of this note – we have attempted some clarification about the use of Cut-off procedure and in applicable approximations in the calculations of ref. [2]. An unambiguous statement about the use of BPHZ regularization procedure is obvious. On the other hand, uses of approximations, however, obvious need caution before use as they sometimes may lead to altogether wrong conclusions.

## V. ACKNOWLEDGEMENT

The authors acknowledge the Financial Assistance from DST (Department of Science & Technology), New Delhi (India) for carrying out this work. P. Parashar wishes to thank CSIR (Council of Scientific & Industrial Research), New Delhi (India) for granting her the Financial Assistance in the form Senior Research Fellow (SRF).

**Table 1:** Theoretical Bounds on $\lambda_W$ and $d_W$ using BPHZ regularization procedure with $m_f \neq m_{f'}$.

| $d_f$ | Experimental limits on $d_f$ $(ecm)$ | Square bracket term $\lvert\chi\rvert$ | $\lvert\lambda_W\rvert$ | $\lvert d_W\rvert$ $(ecm)$ |
|---|---|---|---|---|
| $d_e$ | $\leq 1.8\times10^{-27}$ | $= 4.953\times10^{10}$ | $3.447\times10^{-14}$ | $\leq 4.228\times10^{-30}$ |
| $d_\mu$ | $\leq 3.7\times10^{-19}$ | $= 1.15910^{6}$ | $1.464\times10^{-1}$ | $\leq 1.796\times10^{-17}$ |
| $d_\tau$ | $\leq 3.1\times10^{-16}$ | $= 4.096\times10^{3}$ | $2.065\times10^{1}$ | $\leq 2.563\times10^{-15}$ |
| $d_n$ | $\leq 6.3\times10^{-26}$ (Bare quarks) | $\chi\left\{R' = \left(\frac{m_d}{m_W}\right)^2\right\} = 5.173\times10^{8}$ <br> $\chi\left\{R' = \left(\frac{m_u}{m_W}\right)^2\right\} = 3.593\times10^{8}$ | $1.771\times10^{-11}$ | $\leq 2.172\times10^{-27}$ |

**Table 2:** Theoretical Bounds on $\lambda_W$ and $d_W$ using the expression (4) & (5) for $\Lambda$ = 3 TeV.

| $d_f$ | Experimental limits on $d_f$ $(ecm)$ | $\lvert\lambda_W\rvert$ | $\lvert d_W\rvert$ $(ecm)$ |
|---|---|---|---|
| $d_e$ | $\leq 1.8\times10^{-27}$ | $\leq 7.517\times10^{-32}$ | $\leq 9.222\times10^{-48}$ |
| $d_\mu$ | $\leq 3.7\times10^{-19}$ | $\leq 2.280\times10^{-16}$ | $\leq 2.797\times10^{-32}$ |
| $d_\tau$ | $\leq 3.1\times10^{-16}$ | $\leq 1.241\times10^{-9}$ | $\leq 1.522\times10^{-25}$ |
| $d_n$ | $\leq 6.3\times10^{-26}$ | $\leq 5.825\times10^{-20}$ | $\leq 7.147\times10^{-36}$ |

**Table 3:** Theoretical Bounds on $\lambda_W$ and $d_W$ using the expression (9) of ref. [2].

| $d_f$ | Experimental limits on $d_f$ $(ecm)$ | $\|\lambda_W\|$ | $\|d_W\|$ $(ecm)$ |
|---|---|---|---|
| $d_e$ | $\leq 1.8\times10^{-27}$ | $\leq 1.198\times10^{4}$ | $\leq 1.470\times10^{-19}$ |
| $d_\mu$ | $\leq 3.7\times10^{-19}$ | $\leq 1.191\times10^{3}$ | $\leq 1.461\times10^{-13}$ |
| $d_\tau$ | $\leq 3.1\times10^{-16}$ | $\leq 5.933\times10^{4}$ | $\leq 7.279\times10^{-12}$ |
| $d_n$ | $\leq 6.3\times10^{-26}$ | $\leq 2.052\times10^{-7}$ | $\leq 2.518\times10^{-23}$ |

**Table 4:** Theoretical Bounds on $\lambda_W$ and $d_W$ using the expression (4) & (8) for $\Lambda$ = 3 TeV

| $d_f$ | Experimental limits on $d_f$ $(ecm)$ | $\|\lambda_W\|$ | $\|d_W\|$ $(ecm)$ |
|---|---|---|---|
| $d_e$ | $\leq 1.8\times10^{-27}$ | $\leq 1.385\times10^{-31}$ | $\leq 1.700\times10^{-47}$ |
| $d_\mu$ | $\leq 3.7\times10^{-19}$ | $\leq 3.828\times10^{-16}$ | $\leq 4.697\times10^{-32}$ |
| $d_\tau$ | $\leq 3.1\times10^{-16}$ | $\leq 2.106\times10^{-9}$ | $\leq 2.584\times10^{-25}$ |
| $d_n$ | $\leq 6.3\times10^{-26}$ | $\leq 5.827\times10^{-20}$ | $\leq 7.148\times10^{-36}$ |

**Table 5:** Theoretical EDM Limits on $d_e$, $d_\mu$, $d_\tau$ using $d_W \cong 2.172\times10^{-27}\ ecm$.

| $d_f$ | Theor. Bounds on $d_f$ $(ecm)$ using Eq. (9) of ref. [2] | Theor. Bounds on $d_f$ $(ecm)$ using Eq. (1) | Exp. limits on $d_f$ $(ecm)$ |
|---|---|---|---|
| $d_e$ | $\leq 1.865\times10^{-35}$ | $\leq 9.236\times10^{-25}$ | $\leq 1.8\times10^{-27}$ |
| $d_\mu$ | $\leq 3.855\times10^{-33}$ | $\leq 4.468\times10^{-31}$ | $\leq 3.7\times10^{-19}$ |
| $d_\tau$ | $\leq 6.484\times10^{-32}$ | $\leq 2.656\times10^{-35}$ | $\leq 3.1\times10^{-16}$ |

**Table 6:** Theoretical EDM Limits on $d_e$, $d_\mu$, $d_\tau$, $d_n$ using $d_W \cong 10^{-30}\, ecm$.

| $d_f$ | Theor. Bounds on $d_f\,(ecm)$ using Eq. (9) of ref. [2] | Theor. Bounds on $d_f$ $(ecm)$ using Eq. (1) | Exp. limits on $d_f$ $(ecm)$ |
|---|---|---|---|
| $d_e$ | $\le 8.585\times10^{-39}$ | $\le 4.252\times10^{-28}$ | $\le 1.8\times10^{-27}$ |
| $d_\mu$ | $\le 2.533\times10^{-36}$ | $\le 2.299\times10^{-30}$ | $\le 3.7\times10^{-19}$ |
| $d_\tau$ | $\le 4.259\times10^{-35}$ | $\le 7.224\times10^{-31}$ | $\le 3.1\times10^{-16}$ |
| $d_{n(bare quarks)}$ | $\le 2.251\times10^{-32}$ | $\le 5.722\times10^{-29}$ | $\le 6.3\times10^{-26}$ |
| $d_{n(constituents\ quarks)}$ | $\le 5.263\times10^{-36}$ | $\le 8.270\times10^{-28}$ | $\le 6.3\times10^{-26}$ |

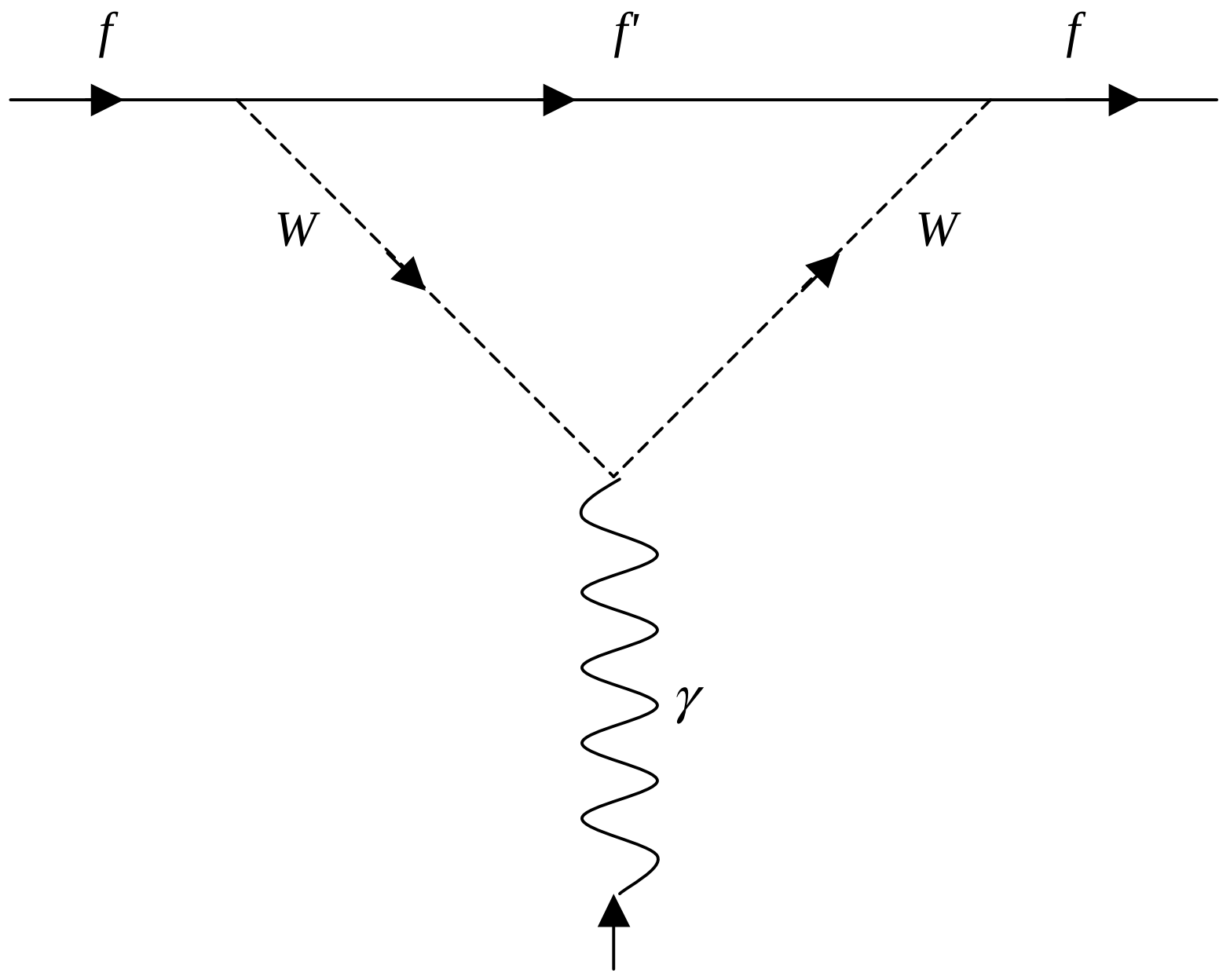

**Fig. 1:** Diagram for the fermion induced electric dipole moment of W-boson.

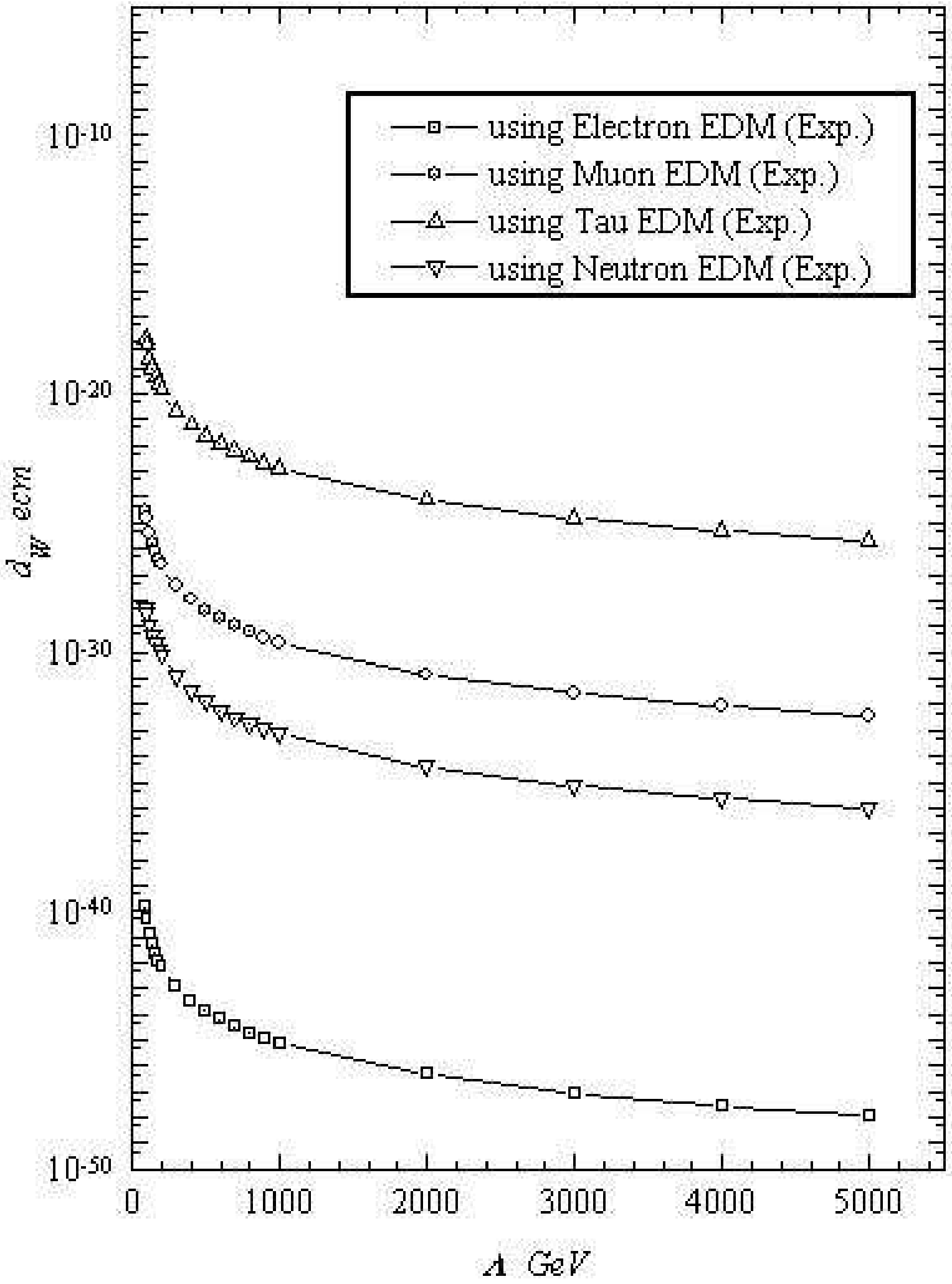

**Fig. 2:** Graph between $d_w$ and $\Lambda$ using our expression (4) & (5).

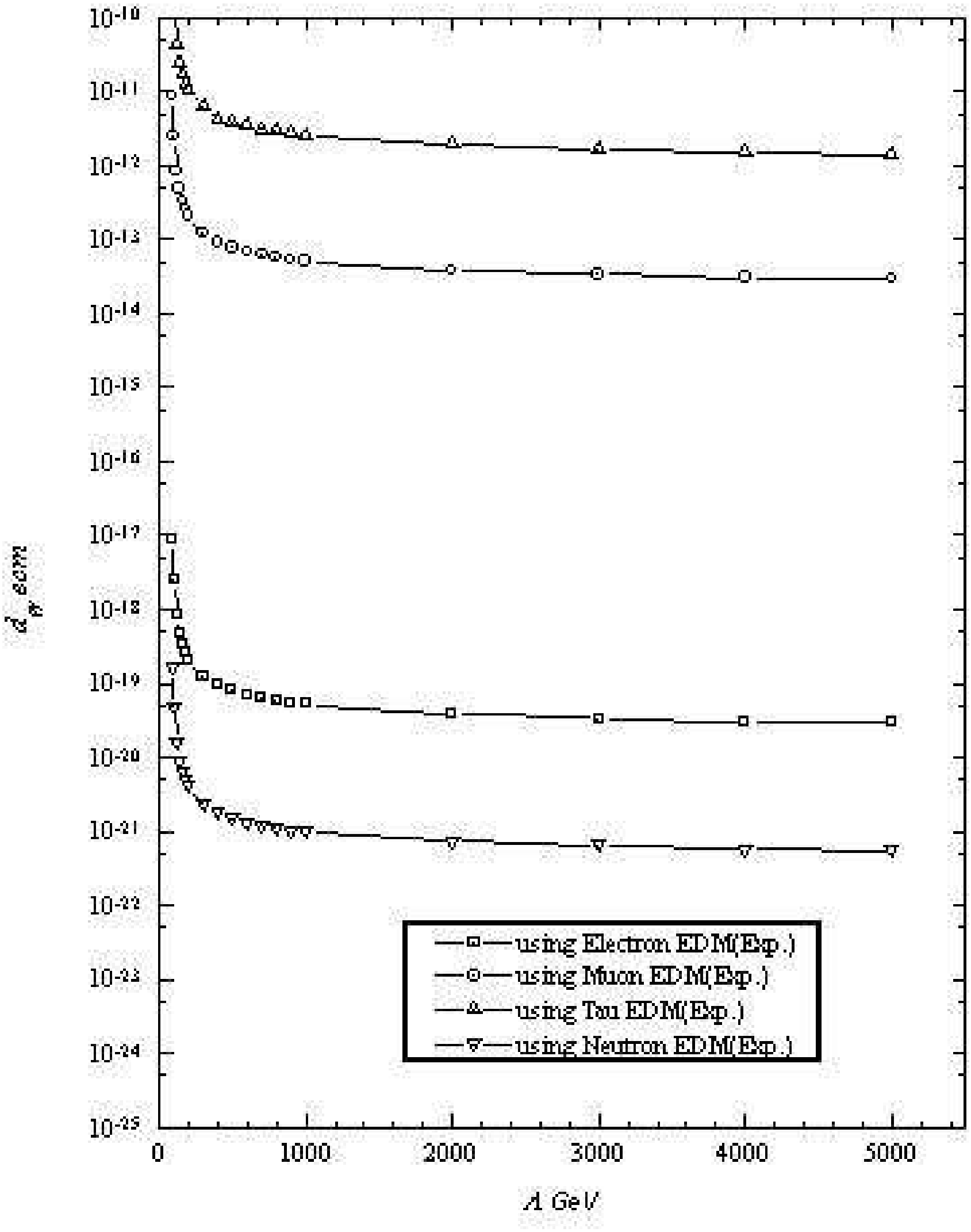

**Fig. 3:** Graph between $d_W$ and $\Lambda$ using expression (9) of ref. [2].

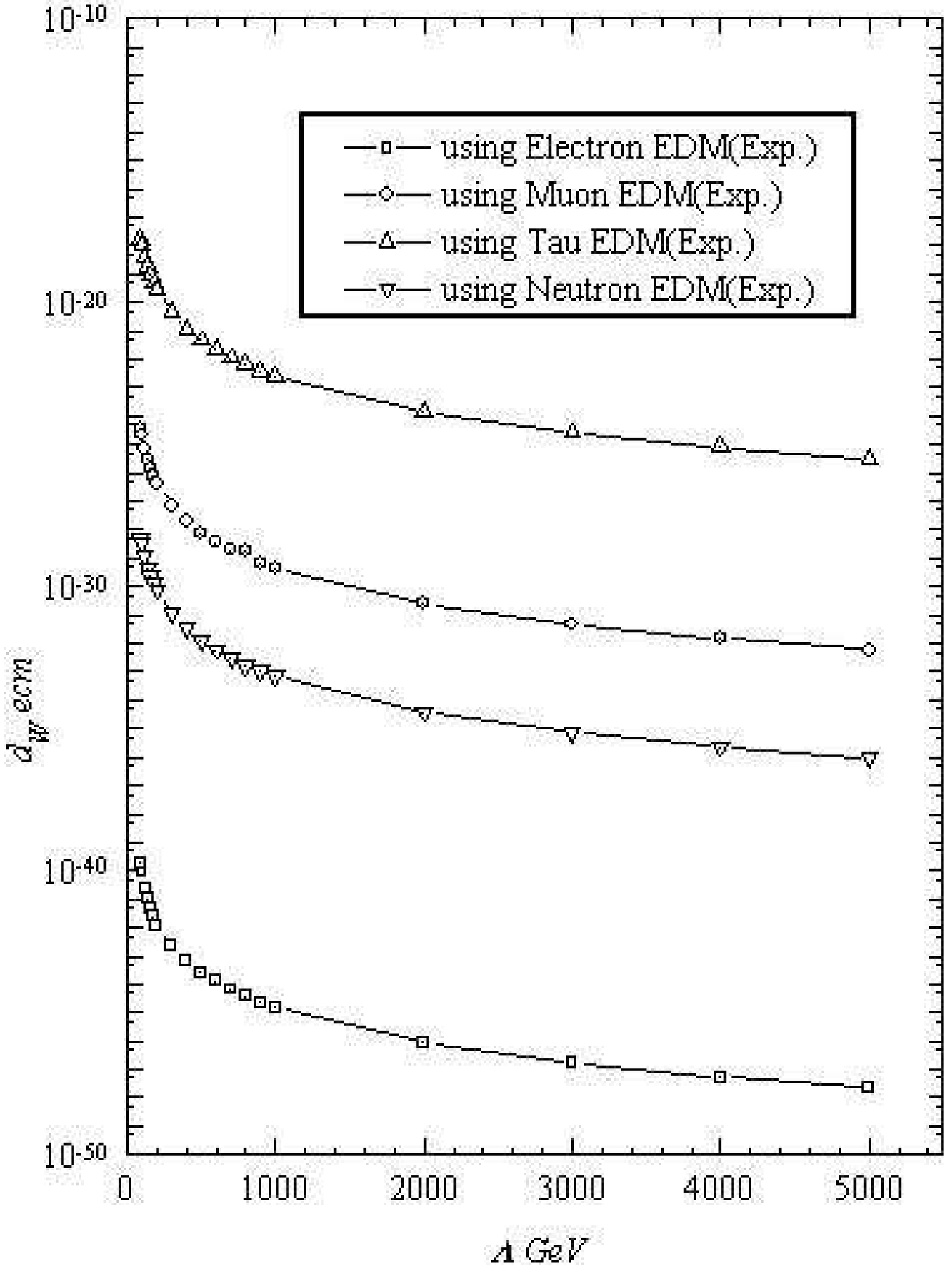

**Fig. 4:** Graph between $d_w$ and $\Lambda$ using our expression (4) & (8).